\documentclass[amsmath,byrevtex,titlepage,
preprint,
aps, superscriptaddress,
nofootinbib ]{revtex4}
\usepackage[T2A]{fontenc}
\usepackage{graphicx}
\usepackage{dcolumn}
\usepackage{bm}
\usepackage{amsmath}
\usepackage{amsfonts}
\usepackage{amssymb}
\usepackage{hyperref}
\usepackage{color}
\newcommand{\beq}{\begin{equation}}
\newcommand{\eeq}{\end{equation}}
\newcommand{\Ro}{{|\cal R}_\omega|^2}
\newcommand{\eq}[1]{eq.~(\ref{#1})}
\newcommand{\Ko}{{|\cal K}_\omega|^2}
\newcommand{\Det}{{\mathfrak Det}_{\rm imp}}
\newcommand{\R}{|R|^2}
\newcommand{\K}{|K|^2}

\newcommand{\la}[1]{\label{#1}}
\begin{document}
\title{
Breaking an one-parameter ``poor man's'' scaling approach in the Luttinger liquid.}
\author{V.V.Afonin}
\affiliation{Ioffe Physical-Technical Institute of the Russian Academy of Sciences, 194021, St.Petersburg, Russia}
\author{V.Yu.Petrov}
\affiliation{
Petersburg Nuclear Physics Institute,  188300, St.Peterscburg, Russia}

\begin{abstract}
Using  derived previously effective theory we explore  conductance in the Luttinger model with one impurity.
 A new approach to the renormalization group (RG) analysis of this  model is developed. It is based on the original Gell-Mann-Low formulation of RG.
 We sum up infrared logarithmic contibutions to conductance in the leading and few subsequent approximations. We analyze the validity of widely used
 ``poor man's'' scaling approach and find that it is applicable only in the leading approximation. Our results for corrections to this approximation
are different from results obtained in other papers.
It should be expected beforehand, as Gell-Mann-Low function of the model is not regularization scheme invariant. For this reason the observed quantity (e.g., conductance) can not satisfy the Gell-Mann-Low equation beyond  the "leading-log"\ approximation as it is supposed in the "poor man's"\ approach.
We formulate the method to calculate the conductance from renormalized hamiltonian in the post-leading approximations and match results to the case of
weak impurity where the answer is known in any order in electron-electron interaction.
\end{abstract}
\maketitle
%
\section{Introduction}

Study of  one-dimensional interacting fermions has a very long history \cite{T}.  Pure 1D systems with arbitrary electron-electron interactions which are described
by the Luttinger model \cite{L} are completely understood both in the fermion and boson language.  Electron transport in these systems is described by ballistic formulae
with   the Fermi speed renormalized by the interaction. A single impurity  doped into an 1D channel  strongly changes the conductance of the channel. Conductance  remains ballistic for low electric field frequency $\omega$ if  the   electron-electron interaction is attractive, but  goes to zero for repulsive interaction.  Conductance exhibits scaling behavior and changes as some power of  frequency for finite $\omega$\cite{FK}.
This behavior is due to infrared divergencies arising for the conductance at small $\omega$.   As a result, summation of an infinite number of perturbative contributions becomes necessary\cite{FK}.

The renormalization group (RG) approach  is a natural way to sum  the infrared logarithms.  The RG methods were applied to calculation of  the conductance ${\cal C}(\omega)$ in the leading logarithmic approximation in \cite{M}. This approximation is valid if
\beq
\nu\log\frac{M}{\omega}\le 1, \qquad \nu \equiv v_c^{-1} -1\ll 1,
\la{applicable}
\eeq
where $M$ is an ultraviolet cutoff   and $v_c$ is the dimensionless renormalized Fermi speed (in the units of the bare one). Commonly used Luttinger parameter is:   ${\tt K}=v_c^{-1}$.

The results  obtained in this approach coincide with the conclusions of Kane and Fischer \cite{FK}   who used the bosonization method.
However, it is  known \cite{gornyi} that the approach suggested in \cite{M} does not work beyond  the leading logarithmic approximation.
Another RG approach was developed in \cite{aristov}  where  the two- and three-loop contributions  were calculated.

The main purpose of this paper is to calculate conductance of the system as a function of the frequency of external electric field (at zero temperature). Most papers dealing with this problem  are using the so called "poor man's"\ RG approach \cite{And}. This approach is also widely used in other problems of condensed matter physics. In the relation to Luttinger liquid in the fermion language "poor man's" RG is based on two main assumptions. First, it is assumed that this theory is determined by effective renormalized Lagrangian which depends only on one "charge". The flow of this charge as a function of external parameters (frequency, temperature, etc) is determined by a Gell-Mann-Low RG equation. Second, it is assumed that physical conductance of the wire is immediately related to this single charge of the theory. Thus conductivity itself should obey RG equation. To find this equation in the given order of electron-electron interaction it is sufficient to calculate all diagrams  and  extract the coefficient in front of the terms with first power of $\log M$ ($M$ is the ultraviolet cutoff). Higher powers of this logarithm should be reproduced by RG equation in the previous order. Then one has to substitute bare conductivity entering extracted coefficient by exact conductivity and include it to RG equation. It is assumed that RG equation obtained this way sums up all infrared logarithms with corresponding accuracy. This procedure was used in \cite{M} in the leading order and in \cite{aristov} in subsequent orders.

Assumptions of poor man's RG were never proved. Moreover, it is clear that it should become wrong in some order of perturbation theory. In fact, no physical quantity can coincide with renormalized charge of the theory or enter renormalized Lagrangian. Therefore they  cannot obey any type of RG equation.

Calculation of physical quantities (e.g. conductivity) in the renormalizable theory always consists of two stages. First, one calculate renormalized effective action defined on some arbitrary scale $\mu$. This can be done with help of Gell-Mann-Low equation which determines renormalized coupling constants. They depend both on the normalization point $\mu$
and regularization scheme of the calculations. On the second stage one is using renormalized perturbation theory in order to calculate conductivity. Renormalized perturbation theory
also depends on $\mu$ and regularization scheme in such a way that dependence on normalization point and regularization scheme cancels. As a result, physical quantity does not depend neither on $\mu$ nor on scheme of regularization. Infrared logarithms can be present or absent in the renormalized perturbation theory depending on the choice of $\mu$. But even if there are no logarithms in renormalized perturbation theory one has to take into account also finite terms. They give contribution in calculations of conductivity beyond the leading logarithmic approximation. Also they are vital to cancel dependence of coupling constants on the regularization scheme. Therefore, poor man's RG should break down in any approximation beyond the leading one.

In \cite{Dual} we formulated another approach to the calculations of the conductance in the Luttinger model with impurity. The original 1+1 fermion theory (one spatial and one time variables) was reduced to an effective boson theory in 0+1 dimensions  (only one variables - the trajectory time) describing the dynamics of the electron phase  at the point of  the  impurity. Effective theory is nonlocal and includes the infinite number of terms describing multiple rescatterings off the impurity. Nevertheless, the effective theory is exactly equivalent to the fermion one. Transition to effective theory simplifies the calculations greatly.

Generally speaking, our approach is a different version of the functional bosonization  used in \cite{Grishin,Mirlin}.  The effective  theory has  another advantages: it is formulated not in terms of the impurity potential but uses the bare reflection ($R$) and transition ($K$) coefficients instead. As a result, the theory  does  not contain additional ultraviolet divergencies appearing due to electron scattering  off the pointlike impurity. There are no divergencies at all if the electron-electron ({\it ee}) potential has a finite radius. This simplifies formulation of the RG method in the effective theory for the point-like impurity.

As shown in \cite{Dual,Dual1} conductance is directly related to the effective reflection coefficient which can be calculated as a Green function of the electron phase in the effective theory. The theories  with electron attraction and repulsion are dual each other: two theories  related by the duality transformation $\Ro\leftrightarrow\Ko$, $v_c\to v_c^{-1}$ are exactly equivalent (see for the details \cite{Dual} and references therein). This theorem allows us to discuss only one of the theories, e.g. attractive one.

The effective theory is derived for an arbitrary {\it ee} potential and pointlike impurity. It is usually assumed that in the region $\omega a\ll 1$ (where $a$ is  the  characteristic scale of e-e interaction) the physical quantities are {\em universal} and independent of the details of the interaction. Then the {\it ee} potential can be replaced by  a point-like $\delta$-functional potential. This  replacement introduces ultraviolet divergences which should be cut off at $M\sim a^{-1}$ where the bare theory is defined. One can determine $M$ comparing physical quantities in the
point-like approximation with results of calculations with real $e-e$ potential.

Conductance is a universal function of the frequency calculable in the RG theory. In the present paper we formulate the Gell-Mann-Low renormalization procedure \cite{GML}  (for  a systematic modern review see, e.g., \cite{collins}) in the effective theory.  Quantum field (phase of electron scattered off the impurity modified by $ee$-interaction, $\alpha (t)$) in the effective theory is dimensionless and  the renormalized Lagrangian contains an infinite number of terms and coupling constants. Nevertheless, the renormalized Lagrangian does not depend on small distances and  the  Lagrangian  reproduces itself after renormalization,  just as it happens in a renormalizable theory.

It is a common wisdom that theories with infinite number of coupling constants are not renormalizable. It is, indeed, true for the theory  with a dimensional fields and therefore with finite number of possible counter-terms of the given dimension. Renormalization of terms of high dimensions leads to power divergencies which depend on the details of the theory at small distances. These corrections destroy the scaling in the theory. However, in our case the role of the field is played by dimensionless electron phase and for this reason there is
 an infinite number of dimensionless counter-terms. All of them are relevant and all ultraviolet divergencies are only logarithmic.

Renormalization group procedure aims to construct the effective action which is free from ultraviolet divergencies and describe the theory at some
arbitrary scale $\mu$. In the Gell-Mann-Low approach it is done by introduction counter-terms in the original bare action. Another idea is used
in the so-called functional renorm group (FRG) (see, e.g., review \cite{FRG}). Here one calculate the effective action  with all fluctuations with momenta larger than $\mu$ integrated out. To do this one modifies electron Green function introducing some artificial cut-off of the small momenta. The effective action on the scale $\mu$ is defined as a Legendre transformation of the bare action with modified electron propagator. This approach  is close in spirit to the Wilson formulation of the renorm group but can be precisely formulated in any number of loops. Moreover, one can derive the infinite system of {\em exact} renorm-group equations. These equations are, in fact, Schwinger-Dyson equations for the given theory. To solve them one has to {\em truncate} this system at some level changing exact vertices to the bare ones.

In principle, results of FRG are equivalent to the Gell-Mann-Low RG with some scheme of regularizaton (the role of scheme is played by the way how electron propagator is modified). However, at the given level of truncation the renormalization flow is not exactly the same: it contains additional terms as compared to Gell-Mann-Low. Strictly speaking, additional terms excess the accuracy and are not valid. Still they believed to give the useful interpolation to the complete RG flow. The back side of FRG method is complexity of equation. Even for the simplest truncation (which corresponds to 1-loop approximation) the complete flow can be found only numerically.

Luttinger liquid with impurity in terms of electrons was considered from this point of view in \cite{fRG}. With the approximation based on the simplest truncation some useful interpolation formula was obtained. Also Authors formulated a hypothesis about behavior of the conductance in the region of very small transition coefficient (for electron attraction). According to this hypothesis the new type of frequency dependence $\omega^{2({\tt K} - 1)/{\tt K}}$. Below (section III) we will obtain analytically conductance in this region and confirm suggestion of \cite{fRG}.

In the paper we calculate all diagrams for the conductance up to the fourth loop, collect infrared logarithms by means of the fourth order Gell-Mann-Low equation, and compare the results with the ``poor man's''\ RG approach. In the leading order approximation the effective reflection (transition) coefficient $\Ro$ is proportional to only one coupling constant and renormalization of this constant does not depend on other coupling constants.  The requirements of  the ``poor man's'' method are satisfied in this approximation, and our approach gives the  same results as in \cite{M} which appears to be  valid in the "leading-log" approximation. In the next-to-leading (two-loop) approximation we collect the logarithmic contributions  of order $\nu^{n+1}\log^n M/\omega$, and as a result more coupling constants enter the Gell-Mann-Low equations. In this and in the three-loop approximation we can diagonalize the system of Gell-Mann-Low equations for the coupling constants by introducing new coupling constants which are functions of the original ones.  However, the ``poor man's'' approach does not work  already starting from two-loop approximation, since the effective reflection coefficient  does not coincide  with  the renormalized charge any more. Therefore, while the two- and three-loop Gell-Mann-Low equations below are the same as in \cite{aristov}, the respective results for $\Ro$ and conductance are completely different. Starting  with the four loops  the ``poor man's'' scaling fails completely.

The second hypothesis of the ``poor man's'' scaling is that observed quantity (conductance) is determined by Gell-Mann-Low equation. This assumption cannot be valid for arbitrary
number of loops. Indeed, it is well-known that beginning from some loop Gell-Mann-Low equations depend on the regularization scheme. Meanwhile, observables cannot depend on the
regularization scheme which is only a specific method of our calculations. For this reason observables cannot enter Gell-Mann-Low equations: these equations are written for
unphysical coupling constants (charges) of the theory. Observables are related to coupling constants  by means of renormalized perturbation theory. Perturbation theory depends
on the regularization scheme as well. This dependence cancels the scheme dependence of the coupling constants.

Usually coupling constants depend on the scheme starting from the third loop, so observables cannot obey RG equations with this accuracy. However, Luttinger model with impurity is a special case: here Gell-Mann-Low eqs depend on the scheme already in the second loop. For this reason we expect that ``poor man's'' approach should break down at the second loop.
We confirm this expectation in Section 3.

Effective reflection coefficient can be calculated in all loops in the limiting case of small reflection (transition) coefficient. One can confront effective reflection (transition)
coefficient in this limit with expressions obtained by RG methods.  Our formulae pass this check while expressions derived from ``poor man's'' RG do not.

The paper is organized as follows. In Section 2 we remind the main points of the effective theory. In Section 3  RG approach to the effective theory is formulated and expressions for conductance in four loops are derived. In Section 4 we match it to the limit of small reflection (transition) coefficient.

In Appendix A we make a connection of our approach with standard one based on Kubo formula. In Appendix B we show explicitly that Gell-Mann-Low function of the model is scheme dependent in two loops. This proves again that ``poor man's`` RG cannot be valid beyond one loop.

\section{Conductance in the effective theory}

 Original Hamiltonian describing interaction of one-dimensional electrons and scattering on the impurity is
\beq
{\cal H}=\int\! dx\;\Psi^+\left[\frac{-\nabla^2}{2m}+V_{\rm imp}(x)\right]\Psi+\int\! dx dy\;\Psi^+\Psi(x)V(x-y)\Psi^+\Psi(y).
\la{HLuttinger}
\eeq
We will assume that the size of the impurity is of order $1/p_F$  ($p_F$ is the Fermi momentum), while the effective radius of the interaction potential $V(x-y)$ is much larger. For this reason electron-electron scattering cannot change the direction of  motion of an electron near the Fermi surface. One can divide electrons into the right  moving (R) and left   moving (L)
\beq
\Psi(x,t) = \psi_R(x,t)e^{ip_F x-i\varepsilon_F t}+\psi_L(x,t)e^{-ip_F x-i\varepsilon_F t}
\eeq
with slowly varying functions $\psi_{R,L}(x,t)$.

Backward scattering is possible only at the  position of the impurity which is usually  modeled by a pointlike potential at $x=0$:
\beq
{\cal H}_{imp}=V_0 \int dx\left[\psi^+_R(x)\psi_L(x)\delta (x)+h.c.\right].
\la{modelRL}
\eeq
Bosonization of this model leads to  the 0+1-dimensional sine-Gordon model considered in \cite{FK}.

Transition to the model \eq{modelRL} of backward scattering induces ultraviolet divergencies. They are due to the fact that in the vicinity of the impurity classification of particles as R- and L-movers is inadequate. As a result, linearized Schr\''odinger equation for $R$- and $L$-particles with potential \eq{modelRL} has no solution. Near impurity one has to return to original equation with full Hamiltonian \eq{HLuttinger}  to find electron wave function. Only after that Hamiltonian can be linearized.

In \cite{Dual} we suggest to describe scattering off the impurity  by means of the bare reflection   and transition coefficients instead of using model (\ref{modelRL}). Electron wave functions obey first order Schr\"odinger equation with {\em linearized} Hamiltonian  and point-like impurity plays a role of boundary condition at $x=0$. In this theory there are no artificial ultraviolet divergencies as in model (\ref{modelRL}). For electrons near the Fermi surface bare transition and reflection coefficients can be replaced by their values $R$ and $K$ at $p_F$.

Besides transition and reflection coefficient electron wave function at the point of impurity depends on some phase which is continuous at $x=0$.
The difference $\alpha$ of the phases of electron wave function for $R$- and $L$-movers is the only dynamical degree of freedom for the problem of interacting electrons (see Ref.\cite{Dual}). One can replace this problem  by a problem of non-interacting fermions in an arbitrary external field  with the help of the well-known Hubbard trick \cite{Hubbard}. Electron transport through the impurity is determined then by the phase $\alpha$
acquired by electron in the Hubbard field.

It is convenient to introduce phase $\alpha$ directly to the path integral determining linear response of the system to the external and integrate out
Hubbard fields. This can be done, as usual, by Faddeev-Popov method \cite{FP}. Integral in Hubbard fields appears to be Gaussian
\footnote{In particular, this is the reason why Luttinger model {\em without} impurity can be solved exactly} and can be calculated exactly.
After that we obtain an effective theory with phase $\alpha$ as the only field. It is the field theory in 1+0 dimension since $\alpha$ depends only on time. Action of the effective theory was derived in \cite{Dual}. It is a sum two terms, the first  describes the electron-electron interaction that is quadratic in  phase
\beq
S_{kin}=\int\!\frac{d\omega}{2\pi}\frac{\alpha(\omega)\alpha(-\omega)}{2W(\omega)}, \qquad
W(\omega)=-\int\frac{dk}{2\pi i}\frac{4k^2V(k)}{(\omega^2-k^2+i\delta)(\omega^2-v_r^2(k)k^2+i\delta)},
\la{kin-energy}
\eeq
where $V(k)$ is the Fourier transformation of the electron-electron potential and $v_r(k)$ is the renormalized Fermi speed
\beq
v_r(k)=\sqrt{1+\frac{V(k)}{\pi}}
\eeq
(we  use the system of  units, where $\hbar$=$v_F=1$).

The second term  in the action  describes multiple rescattering  off the impurity:
\begin{eqnarray}
S_{int}[\alpha]\!&=&\!-\log\Det[\alpha]\!=\!\sum_{n=1}^\infty \frac{(-1)^{n}}{n}\!\left(\!\frac{|R|}{|K|}\!\right)^{2n}\!{\cal B}_{2n-1}[\alpha]
,\nonumber\\
{\cal B}_{n}\!&=&\!\int\!\!\frac{d\tau_0\ldots d\tau_n}{(2\pi i)^{n+1}}
\frac{1\!-\!\cos[\alpha(\tau_0)\!-\!\alpha(\tau_1)\!+\!\ldots \alpha(\tau_n)]}
{(\tau_0-\tau_1-i\delta)\ldots(\tau_n-\tau_0-i\delta)}.
\la{det1}
\end{eqnarray}
It plays the role of interaction in the effective theory. Formally, this term appears as a result of integration in fermion fields and represents a part of fermion determinant related to impurity.

Effective theory with the action $S=S_{kin}+S_{int}$ is exactly equivalent to the original Luttinger model with an impurity.
The physical observables  in the Luttinger model should be calculated as averages
\beq
\langle\langle \ldots \rangle\rangle = \frac{1}{{\cal Z}}\int\! D\alpha(\tau) \ldots e^{-S_{kin}[\alpha]}{\Det}[\alpha].
\la{boson}
\eeq

Our primary goal now is calculation of the conductance ${\cal C}(\omega)$. At low frequencies $\omega a\ll 1$ ($a$ is the characteristic size of the electron-electron potential)
conductance is expected to be a universal function independent of the details of the potential \cite{FK}.
 Then we can  replace it by a pointlike potential
\beq
V(x-y)=V_{ee}\delta(x-y).
\eeq
The quadratic form $W(\omega)$ for this potential turns into
\beq
W(\omega) = \frac{2\pi}{|\omega|}\left[\frac{1}{v_c} -1\right]\equiv \frac{2\pi\nu}{|\omega|} .
\label{W}
\eeq
 Here $v_c=\sqrt{1+{V_{ee}}/{\pi}}$ is a renormalized Fermi speed for the short-range  e-e interaction and $\nu$ can serve as measure of the strength of the electron-electron interaction. Theories with $V_{ee}<0$ (electron-electron attraction) and $V_{ee}>0$ (electron-electron repulsion) are {\em dual} each other \cite{FK, Dual}. For this reason we consider only the theory of  mutually  attracting   one-component (i.e. without spin) electrons and calculate the effective
reflection coefficient $\Ro$.

The effective reflection coefficient in the effective theory  in terms of the Green function of the $\alpha$-field  has the form
\beq
\Ro=\frac{\nu+1
}{\nu^2}\frac{|\omega|}{2\pi}[G_0(\omega)-G(\omega)],
\la{obs2}
\eeq
where $G(\tau)=\langle\langle\alpha(\tau)\alpha(0)\rangle\rangle$ and $G_0(\omega)$ is  the free Green function.

The reflection coefficient is a physical observable, conductance ${\cal C}(\omega)$ is simply related to $\Ro$:
\beq
{\cal C}(\omega)=\frac{e^2\Ko}{2\pi v_c}, \qquad \Ko=1-\Ro.
\la{obs14}
\eeq

Formulae for conductance, \eq{obs2}, \eq{obs14} look a bit unusual. For this reason we explain in Appendix A their relation to the familiar Kubo formula.
The relation is based on well-known Ward identities .

Using the effective theory \eq{boson} build in Ref.\cite{Dual} and \cite{Dual1} we can develop perturbation theory in  electron-electron interaction for the reflection coefficient
$\Ro$. This theory is far simpler than in the original Luttinger model, as the effective theory is only 0+1-dimensional. Perturbation theory corresponds to the small electron phases $\alpha$. Expanding $\Det$  in \eq{boson} in $\alpha$ (we use \eq{det1}) we obtain $\Ro$ as a sum of Feynman diagrams. Propagators in these diagrams are $W(\omega)$, \eq{W}, and
vertices should be extracted from \eq{det1}. Vertices with any number of legs are possible. According to \eq{obs2} reflection coefficient is a sum of diagrams with two external legs.

It is well-known \cite{FK,M} that perturbation theory in Luttinger model contains logarithms of the frequency $\omega$ of the external electrical fields. At small $\omega$ it is necessary to sum them up. For this purpose we will use renormalization group (RG) methods.


\section{Renormalization group approach}
\label{RGcalcul}

We  use the RG method  to sum the  logarithmically divergent corrections to the reflection coefficient. In  the original formulation  of the renormalization group \cite{GML,collins} one has to find counterterms compensating all ultraviolet divergencies and include them into the renormalized action
\begin{eqnarray}
S_r[\alpha]&=&\sum_{n=1}\frac{g_{2n}(\mu )}{4\pi}\int\frac{d\omega_1...d\omega_{2n}}{(2\pi)^{2n}(2n)!}\Gamma_{2n} (\omega_1...\omega_{2n})\cdot \label{Gama}
\nonumber\\
&&\cdot\alpha (\omega_1)...\alpha (\omega_{2n})\delta (\omega_1+...\omega_{2n}),
\end{eqnarray}
Here $g_{2n}(\mu)$ are  the renormalized coupling constants\footnote{We are using a bit non-standard version of the Gell-Mann-Low theory.  Besides other coupling constants we introduce  also the coupling constant $g_{2}$ for the vertices with only two legs. Respectively, there is no need in the $Z$-factor that renormalizes field $\alpha(\omega)$.}
\beq
g_{2n}(\mu) = g^{(0)}_{2n}+\delta g_{2n}(\mu),
\eeq
where  $g^{(0)}_{2n}$  are bare coupling constant and $\delta g_{2n}(\mu)$  are sums of counterterms normalized at some point $\mu$. In what follows, we will
use  the Pauli-Villars regularization. The physical quantities (in particular,  the reflection coefficient)  do not depend on  the regularization scheme.  The renormalized coupling constants should be normalized by  the condition $g_{2n}(\mu\!=\! M)=1$, what means that at $\mu=M$  the renormalized Lagrangian coincides with  the original one in \eq{det1}.

 For dimensional reasons the vertices $\Gamma_n$ should be linear in  frequency.  Also  they should nullify if any of the frequencies $\omega_i=0$.  Duality between the theories with electron-electron attraction and repulsion  imposes very strong restrictions on the vertices $\Gamma_n$.  These requirements determine the vertices up to a constant
\beq
\Gamma_{2n} (\omega_1,\ldots ,\omega_{2n})\!=\!\left[ S_n\theta\left(\prod_i\omega_i\right) +A_n\theta\left(-\prod_i\omega_i\right)\right]\gamma_{2n}(\omega_1,\ldots,\omega_{2n}),
\eeq
where $\gamma_{2n}$  are functions of the external frequencies,
\beq
 \gamma_{2n} (\omega_1,\ldots,\omega_{2n})=\sum_i|\omega_i|-\sum_{i<j}|\omega_i+
\omega_j|+\sum_{i<j<k}|\omega_i+\omega_j+\omega_k|-...,
\eeq
and $S_n$, $A_n$ are  functions of the bare reflection ($\R$) and transition ($\K$) coefficients, $S_n$ being symmetric and $A_n$ antisymmetric under the $\R\leftrightarrow\K$ exchange. The functions $S_n$ and $A_n$  are subject to the  recurrence relations
\beq
S_n(x)=-\frac{\partial A_{n-1}(x)}{\partial \log{x }},
\quad
A_n(x)=-\frac{\partial S_{n-1}(x)}{\partial \log{x }},
\eeq
and
$$
S_1=\frac{x}{1+x}, \quad S_2=-\frac{\partial S_1}{\partial \log{x }}=-S_3, \quad A_1=A_2=0,
$$
where $x=|R|^2/|K|^2$. Respectively, for any  $n>2$ there are two constants $g^{(s)}_{2n}$ and $g^{(a)}_{2n}$ in front of the symmetric and antisymmetric structures which should be renormalized independently.

\begin{figure}
\begin{center}
\includegraphics[width=10 cm]{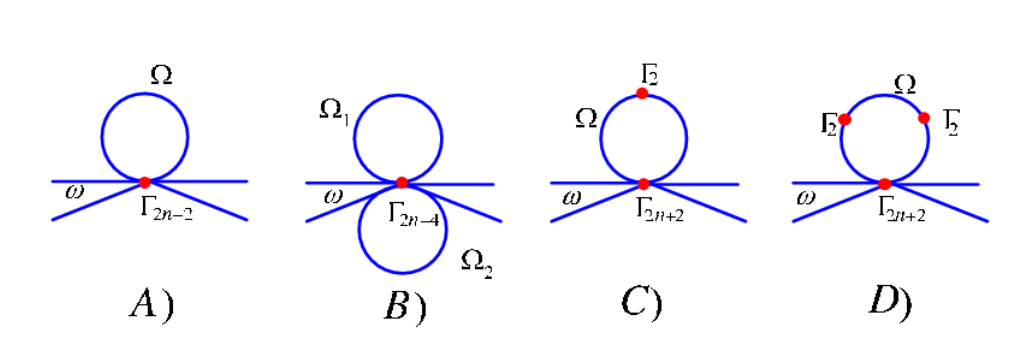}
\caption{ Diagrams contributing to renormalization of coupling constant $g_{2n}$: A)---order $\nu$;
B),C)---order $\nu^2$; D)---order $\nu^3$}
\la{fig1}
\end{center}
\end{figure}

Now let us proceed with the renormalization program.
Only the diagram in Fig.1A renormalizes  vertex $\Gamma_{2n}$ in the first order in $\nu\ll 1$. Explicitly
\beq
\{1A\} =\frac{g_{2n+2}}{4\pi}\int\frac{d\Omega}{2\pi}G_{P.V.}(\Omega)\Gamma_{2n+2}(\Omega,-\Omega,\omega_1,\ldots,\omega_{2n})
 \alpha(\omega_1)\ldots\alpha(\omega_{2n}),
\la{1a}
\eeq
where
\beq
G_{P.V.}(\omega)=\frac{2\pi\nu M_{P.V.}}{|\omega|(|\omega|+M_{P.V.})},
\label{Gpv}
\eeq
is  the bare Green function regularized according to Pauli-Villars. The integral in \eq{1a} is logarithmically divergent for $|\Omega|\gg |\omega|$.  To compensate this divergence one has to add the counterterm
\beq
\delta g_{2n}^{(s)}(\mu )=-2\nu g_{2n+2}^{(a)}(\mu) \log\left(\frac{M_{PV}}{\mu}\right)\frac{1}{S_{n}}
\frac{\partial S_n}{\partial \log{x }},
\label{Z2}
\eeq
to the action,
and  also add a similar counterterm  $\delta g_{2n}^{(a)}(\mu )$ with the substitution $S\to A$.

Three diagrams contribute to renormalization in the second order in $\nu$.  These are the diagrams 1B, 1C, and the diagram 1A with the vertex replaced by the counterterm in \eq{Z2}.
The sum of 1A and 1B  could contain  divergencies proportional to $\log^2 M_{P.V.}$ and $\log M_{P.V.}$, but  linear logarithm does not arise in our regularization.
Let us emphasize that the coefficients before the  divergent  terms do not depend on external frequencies $\omega_i$.  Therefore, the divergencies can be compensated by {\em local} counterterms:
\begin{equation}
\delta g_{2n}^{(s)}[IB\!+\!IA]=-2\nu^2 g_{2n+4}^{(s)} \log^2{\left(\frac{M_{PV}}{\mu}\right)}\frac{1}{S_{n}}
\frac{\partial^2 S_n}{\partial \log{x }^2},
\label{nu2}
\end{equation}
as it should be in a renormalizable theory. Moreover,  the counterterm  in \eq{nu2} does not contribute to the $\beta$-function. Its contribution is canceled by the contribution of the
first order counterterm \eq{Z2} containing the derivative $dg_{2n+2}(\mu)/d\mu$ that should be accounted for in this order.

Only the essentially new diagram  in Fig. 1C contributes to  the $\beta$-function. This
diagram is proportional to the first power of the logarithm and  the corresponding counterterm differs from  the one in \eq{Z2} only by  the factor $-\nu g_2 S_1=-\nu g_2\R $.

To obtain a system of the Gell-Mann-Low equations we collect all counterterms and differentiate them in $\log\mu$
\beq
\frac{\partial g_{2n}^{(s)}(\mu ) }{\partial \log{\mu}} =
\frac{2\nu g_{2n+2}^{(a)}(\mu)}{1+\nu g_2(\mu) \R }
\frac{\partial \log S_n}{\partial \log{x }},
\label{GL1}\eeq
and similarly for $g_{2n}^{(a)}(\mu )$.
These equations account for the sum of diagrams with an arbitrary number of the vertex $\Gamma_2$ insertions in the one-loop  diagrams. Among them is the diagram in  Fig.1C of order $\nu^2$, the diagram  in Fig.1D of order $\nu^3$ and so on.  These are the only diagrams that give contributions  of the respective orders to the $\beta$-function. Only the corrections of order $\nu^4$ would modify the r.h.s. in \eq{GL1} further.

The system of the Gell-Mann-Low equations in \eq{GL1} can be  considered  as a system of recurrence relations  for the coupling constants $g_{2n+2} (\mu )$ in terms of the derivatives of  $g_{2n}$. The first coupling constant $g_2(\mu)$ is not determined by these recurrence relations. Nevertheless,  the boundary conditions  $g_{2n}(\mu\!\!=\!\! M)=1$ for all $n$ are sufficient to find the running coupling constant $g_2(\mu)$ at all $\mu$.

 It is possible to derive a closed equation for $g_2$  with the accuracy used in \eq{GL1}. Dividing  each equation in the system  in \eq{GL1}  by the first  one we obtain for all $n>1$
\beq
h_4\left(\frac{\partial S_1(z)}{\partial z}\right)^{-1}\frac{\partial h_{2n}}{\partial z}=h_{2n+2}, \qquad h_2=\frac{z}{1+z}.
\la{eqforh}
\eeq
We introduced here the notation
\begin{eqnarray}
h_{2n} &=&\left\{S_1g_2,\frac{\partial S_1}{\partial\log x}g_4,\frac{\partial^2 S_1}{(\partial\log x)^2}g^{(a)}_6,\frac{\partial^3 S_1}{(\partial\log x)^3}g^{(s)}_8\ldots\right\},\nonumber\\
f_{2n} &=&\left\{\frac{\partial S_1}{\partial\log x}g^{(s)}_6,\frac{\partial^2 S_1}{(\partial\log x)^2}g^{(a)}_8,\frac{\partial^3 S_1}{(\partial\log x)^3}g^{(s)}_{10}\ldots\right\}.
\la{notation}
\end{eqnarray}
Equations for $f_{2n}$ look similarly.

It is easy to find the solution of \eq{eqforh}:
\beq
h_{2n} = \frac{\partial^{n-1}S_1(z)}{(\partial\log z)^{n-1}}=f_{2n+2}, \qquad
n>1.
\la{solh}
\eeq
Indeed, as can be seen from \eq{notation}, the boundary condition for $g_2$ ($g_2(M)=1$) means that $z=x$ at $\mu=M$.  The boundary conditions for the other coupling constants ($g_{2n}(M)=1$) are also  satisfied at $z=x$, as can be seen from \eq{solh} and the definitions  in \eq{notation}. At last,  the expression  in \eq{solh} explicitly  satisfies \eq{eqforh}.

The solution in eq. (\ref{solh}) allows us to find  the dependence of the coupling constant  $h_4$ on $h_2$ and hence write the first equation of the system (\ref{GL1}) as a nonlinear equation for $h_2$
\beq
\frac{\partial h_2}{\partial\log\mu}= \frac{2\nu}{1+\nu h_2}\frac{\partial S_1(z)}{\partial\log z}=\frac{2\nu}{1+\nu h_2}h_2(1-h_2)
\la{eqh2}
\eeq
This equation determines dependence of $h_2$ on the normalization point.

 In terms of $z(\mu)$ the differential equation in \eq{eqh2} reduces to an algebraic one
\beq
z(\mu)=x\left(\frac{\mu}{M}\right)^{2\nu}\left[ \K(1+z(\mu))\right]
^{-\nu}.
\la{z-equat}
\eeq
 Iterating this equation one can find the renormalized coupling constants in a few lowest orders:
\begin{eqnarray}
g_2(\mu)&=& \frac{1}{\R+\K(M/\mu)^{2\nu}}- \nonumber\\
&&-\nu\K(M/\mu)^{2\nu}\frac{\log[\K+\R(\mu/M)^{2\nu}]}
{[\R+\K(M/\mu)^{2\nu}]
^2}+\ldots,
\nonumber\\
g_4(\mu)&=&\frac{(M/\mu)^{2\nu}}{[\R+\K(M/\mu)^{2\nu}]^2}
+\ldots,
\label{h4}
\end{eqnarray}
and so on.

Let us notice now that  this solution of  the Gell-Mann-Low equation literally coincides with the one in the ``poor man's'' RG approach. Indeed, the coupling constant $h_2(\mu)$ at $\mu=M$ coincides with  the  bare reflection coefficient squared, $h_2(M)=S_1(x)=\R$. At smaller $\mu$  the Gell-Mann-Low function (r.h.s. of \eq{eqh2}) is determined by  the perturbative single-logarithmic terms,  where  one has to substitute $\R\to h_2(\mu)$ (i.e. $x\to z$). This procedure is literally  the ``poor man's'' one, {\em if one identifies}\/ $\Ro=h_2(\omega)$. For this reason \eq{eqh2} exactly coincides with  the Gell-Mann-Low equations derived in \cite{M} in the leading order, and in \cite{aristov} in order $\nu^3$, where this method was used.

 The ``poor man's'' method holds in our problem if the higher order contributions  arise in the Gell-Mann-Low equations \eq{GL1}  as a universal factor that depends only on $g_2$ on the r.h.s. in all equations.
It is this property that allows us to derive the closed nonlinear equation \eq{eqh2} for $h_2$.
This property breaks down already at  order $\nu^4$.
\begin{figure}
\begin{center}
\includegraphics[width=4.5 cm]{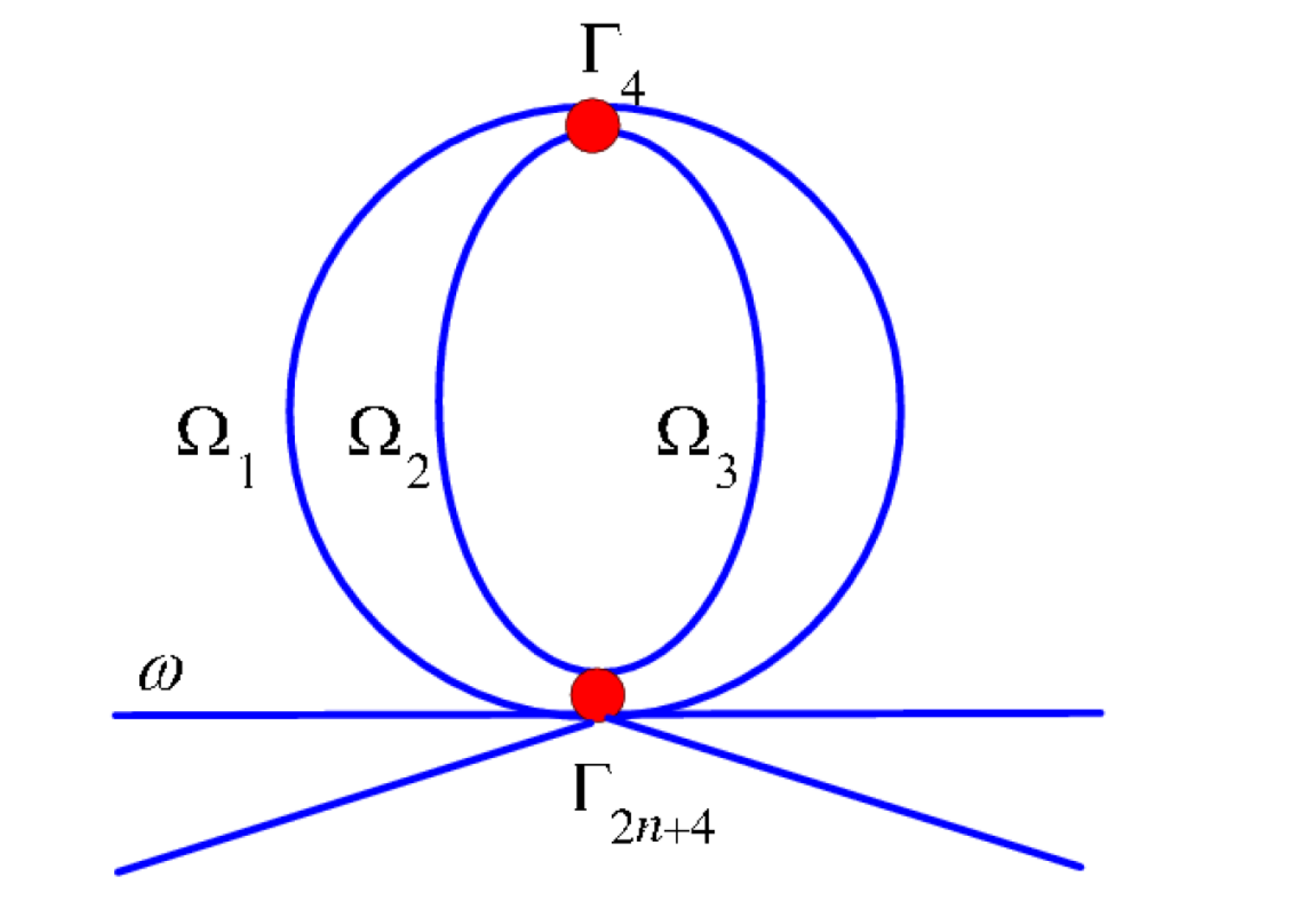}
\caption{ ``Water melon'' diagram $O(\nu^4)$, the first contribution to the charges not described
by ``poor man's''  approach}
\la{fig2}
\end{center}
\end{figure}
 The ``water melon'' diagram in Fig. \ref{fig2}  contains a single-logarithmic  divergence  $\sim\nu^4\log M_{PV}$ and  therefore  generates an additional contribution to the r.h.s. in \eq{eqforh}
 \beq
\frac{\partial h_{2n}}{\partial\log\mu}=\frac{2\nu h_{2n+2}}{1+\nu h_2}-\frac{\pi^2}{3}\nu^4h_4 h_{2n+4}.
\la{delta1}
\eeq
It is the only diagram of order $\nu^4$  that modifies the Gell-Mann-Low equation.  We consider the new contribution on the r.h.s in \eq{delta1} as a new source for $h_{2n}$ and linearize \eq{delta1}
\beq
\frac{\partial \delta h_{2n}}{\partial\log\mu}-2\nu\delta h_{2n+2}=-\frac{\pi^2}{3}\nu^4 h_4h_{2n+4},
\la{delta2}
\eeq
where $\delta h_{2n}$ is a small correction to $h_{2n}$. The r.h.s.  should be calculated in the leading order. The solution of this system  is:
\beq
\delta h_2=-\frac{\pi^2}{3}\nu^3\frac{\partial^2 h_2}{(\partial \log{(\mu^{2\nu})^2}}[h_2(\mu)-h_2(M)].
\la{delta3}
\eeq
Meanwhile, the ``poor man's''  recipe implies that we calculate the r.h.s. of \eq{delta1} in the leading order and substitute  $h_2$ instead of each $\R$. This  would lead to  the equation
\[
\frac{\partial\delta h_2}{\partial\log \mu}-2\nu\delta h_2 = -\frac{\pi^2}{3}\nu^4 h_2(1-h_2)(h_2-3h_2^2+2h_2^3),
\]
which is different from \eq{delta1} and has a solution different from  the one in \eq{delta2}. Hence  the  ``poor man's'' RG   fails at order $\nu^4$ and one is left with  a complete system of  the Gell-Mann-Low equations accounting for  an infinite number of  coupling constants.

\begin{figure}
\begin{center}
\includegraphics[width=12 cm]{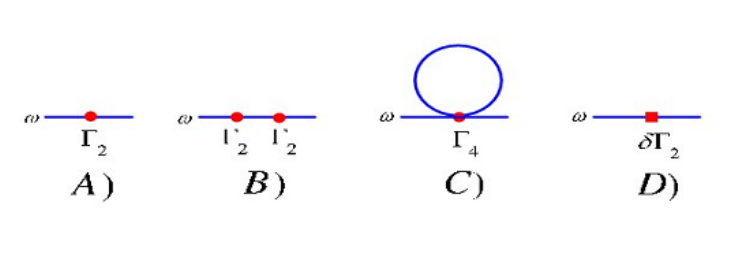}
\caption{ Diagrams contributing to the Green function: A) $\nu$-order; B), C), and D) $\nu^2$ order.
Diagram D) is the contribution of counter-term to $\Gamma_2$ vertex}
\la{fig3}
\end{center}
\end{figure}

Let us turn now to the discussion of the effective reflection coefficient $\Ro$.  We use renormalized perturbation theory and relate $\Ro$ to the Green function by means of \eq{obs2}. Owing to counter-terms there are no divergencies in the renormalized theory. The lowest order diagrams for  the Green function are presented  in Fig.\ref{fig3}. The vertexes $\Gamma_{2n}$ include also counter-terms, coupling constants $g_{2n}(\mu)$ are renormalized ones. It is important that diagrams should be calculated exactly,  including the finite terms (not only divergent parts or logarithmic contributions  as for  the Gell-Mann-Low function). We will see that finite contributions determine coefficients in front of the subleading infrared logarithms.
\beq
\Ro=1 - \nu g_2(\mu)|R|^2+\nu^2 g_2^2(\mu)|R|^4
+2\nu^2g_4(\mu)|R|^2|K|^2(\log{\frac{\omega}{\mu}} + 1) ]+ \ldots
\label{Gz}
\eeq
We have calculated also all diagrams in the renormalized theory up to the fourth order in $\nu$. Thus we are able to give expressions for $\Ro$ in 4 loops but this expression is too cumbersome and will be published elsewhere.
Higher diagrams contain higher powers of $\log\omega/\mu$.

Effective reflection coefficient is a physical quantity and cannot depend on the normalization point  $\mu$. Parameter $\mu$ is an artificial parameter introduced to the theory in order to divide the contribution of high and low
frequencies. First are included into the renormalized coupling constants while contribution of second appeared explicitly in the renormalized perturbation theory. Let us note that separation of frequencies in the Gell-Mann-Low approach is achieved by introducing corresponding counter-terms not like in Wilson approach where momenta integrals are divided in some regions of integration. It is not easy to define Wilson procedure beyond one loop while an approach based on counter-terms is valid in any number of loops.

Independence of $\Ro$ on $\mu$ leads to well-known Callan-Symanczik equation
\beq
\left[\mu\frac{\partial}{\partial\mu}+\sum_{n}\beta_{2n}(g_{i})\frac{\partial}{\partial g_{2n}}\right]\Ro =0
\la{CalSym}
\eeq
Here $\beta_{2n}(g)$ are Gell-Mann-Low $\beta$-functions. Usually Callan-Symanzik equation includes also anomalous dimensions. As we introduced $g_2$ coupling constant, we do not need them (see footnote above).
Effective reflection coefficient is dimensionless, therefore explicit dependence $\Ro$ on $\mu$ can be only in combination $\mu/\omega$. For this reason $\mu\partial_\mu\Ro=-\omega\partial_\omega\Ro$ and
Callan-Symazik equation can be used to determine dependence of $\Ro$ on the frequency. The number of terms which should be accounted in \eq{CalSym} depends on the relation between $\mu$ and $\omega$.
For the region  $\mu/\omega\gg 1$  logarithmic contributions are present in perturbation theory and one has to sum up all order of renormalized perturbation theory. If $\mu\sim\omega$ we can restrict
ourselves by few lowest orders.

In fact, there is no need to solve \eq{CalSym}. Let us choose $\mu=\omega$. Then there is no logarithmical contributions in renormalized perturbation theory. Diagrams of perturbation theory depend on the
frequency only through coupling constants $g_{2n}(\omega)$. Coefficients in front of the coupling constants are numbers which can depend only on $R$, $K$ --- bare transition and reflection contribution.
All infrared logarithms in this case are included in the coupling constants and hence can be found by renorm-group methods using Gell-Mann-Low equations.

Using normalization point $\mu=\omega$ we arrive at the following expression for effective reflection coefficient
\beq
\Ro = (1+\nu)\R\left[g_2(\omega)-\nu \R g_2^2(\omega)-2\nu g_4(\omega)\K\right].
\la{2loop-all}
\eeq
Only finite contributions to the diagrams, with no logarithms at all, survive here.

Importance of the finite contributions to the diagram  is clear from the following considerations. Let us take the diagram which produces in the leading approximation $\nu^n\log^n(M/\omega)$. Typically such a  diagram consists of the number of sub-diagrams each contributing $\nu\log(\omega/M)$. If we account for the finite contribution (no logarithm) in one of the sub-diagrams we will loose one logarithm and obtain contribution of order $\nu^{n}\log^{n-1}(M/\omega)$. This contribution should be included to the next subleading approximation (next loop). Eq. (\ref{2loop-all}) is sufficient to calculate $\Ro$ in 2 loops (leading and next-to-leading logarithmic approximation) provided that coupling constants are known in the same approximation. As was said, we calculated $\Ro$ also in two next approximations (i.e. calculate all diagrams up to the order $\nu^4$ and coupling constants in four loops).

Physical observable $\Ro$ cannot depend also on {\em regularization scheme}. Meanwhile,
starting from the second loop (terms $O(\nu^2$)) the expressions for the r.h.s. of the Gell-Mann-Low equations (15) and (22) depend on the regularization scheme.
Coupling constants are different in the different schemes  but they are related by redefinition of the cut-off $M$ with a constant factor which can be determined from one-loop diagrams. Dependence on the regularization scheme typically starts from the third loop. Unusual dependence can be traced to the fact that in our theory  expansion of the $\beta$-function starts with the linear and not cubic terms in the coupling constants (we explain this in more details in Appendix B).

The expression \eq{2loop-all} for $\Ro$ depends on the scheme as well, here the coefficient in front of $g_4(\omega)$ is scheme-dependent. This contribution comes from
the diagram 3c. This diagram is logarithmically divergent. Divergency is canceled by the counter-term but finite term depends on the regularization scheme explicitly.
Nevertheless, the final expression for the effective reflection coefficient $R_\omega^2$ {\em does not depend on the scheme} as it should be.
The explicit dependence on the scheme of \eq{2loop-all} is compensated by the dependence on the scheme of the coupling constants (see also Appendix B).

Substituting to \eq{2loop-all} expression for coupling constants we obtain finally
\beq
\Ro =\frac{|R|^2\left(\frac{\omega}{M}\right)^{2\nu}}{|K|^2\!+\!|R|^2\left(\frac{\omega}{M}\right)^{2\nu}}
\left\{1- \frac{\nu |K|^2}{|K|^2\!+\!|R|^2\left(\frac{\omega}{M}\right)^{2\nu}}
\left[1+\log\left(|K|^2\!+\!|R|^2\left(\frac{\omega}{M}\right)^{2\nu}
\right)\right]
\right\}
+O(\nu^2)
\la{final}
\eeq
We have calculated above  contributions to the coupling constants  of order $\nu^4$, what allows us to obtain $\Ro$ with the same accuracy.
We  derived an expression for $\Ro$   with this accuracy,  but it is too cumbersome to be presented here and will be published elsewhere.

Expression (\ref{final}) is valid in the region
\beq
\left(\frac{\omega}{M}\right)^{2\nu} \sim 1
\la{region}
\eeq
All infrared logs  $\nu^n\log^n(M/\omega)$ and $\nu^{n+1}\log^n(M/\omega)$ are summed by the system of  the Gell-Mann-Low equations.

As seen from \eq{2loop-all} identification  of $\Ro$ with the coupling constant $h_2(\omega)=\R g_2(\omega)$ is valid only in the leading order. In the next order this  assumption of  the ``poor man's'' RG breaks down and  the effective reflection coefficient depends in some way on {\em other} coupling constants  in the theory. Correspondingly, \eq{final} in the leading order coincides with expression obtained in \cite{M} but differs from the reflection coefficient derived in \cite{aristov}. Comparison of these results is presented on Fig.\ref{fig4}.
\begin{figure}
\begin{center}
\includegraphics[width=18 cm]{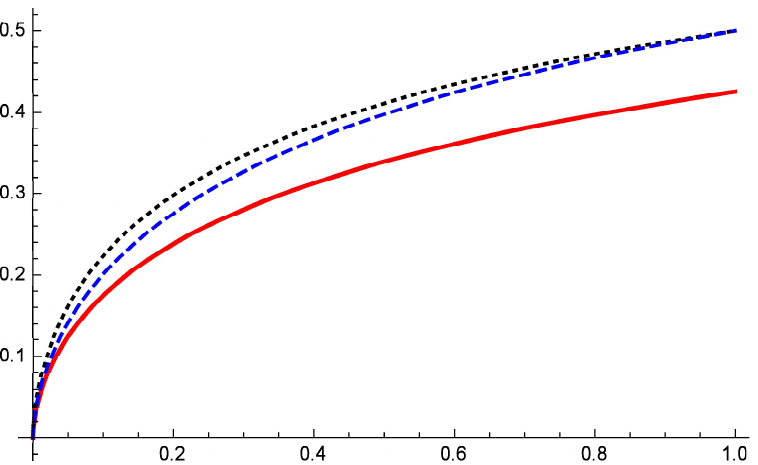}
\caption{Dependence of effective reflection coefficient on the frequency $\omega/M$ at $\nu=0.3$, $|K|^2=|R|^2=\frac{1}{2}$. Dashed line --- 1-loop  approximation of \cite{M},
dotted line --- ``poor man's'' RG \cite{aristov}, solid line --- \eq{final}}
\la{fig4}
\end{center}
\end{figure}

It is clear that effective reflection coefficient $\Ro$ cannot obey any kind of Gell-Mann-Low equations. Indeed, these equations always depend on the regularization scheme while $\Ro$ should not. Hence, in some order
Gell-Mann-Low eqs written for $\Ro$ should fail. In our model it happens in the next-to-leading order, precisely at the moment when Gell-Mann-Low function starts to depend on the regularization. Meanwhile, the idea that to
determine $\Ro$ one has to write RG equation is in the heart of the ``poor man's'' approach. Therefore, this approach should break down in some order of perturbation theory. We see this explicitly in the second loop.

``Poor man's RG'' never takes into account non-logarithmic contributions to the diagrams. However, these contibutions are important starting from second loop. It is assumed in this approach that at $\omega=M$
always $\Ro=|R|^2$. Meanwhile, we obtain from \eq{final} at $\omega=M$
\beq
\Ro = |R|^2(1-|K|^2\nu +O(\nu)^2)
\eeq
The second term here is a finite correction to bare reflection coefficient as due to electron-electron interaction. One can say that this correction is not interesting or even call $|R|^2$ the reflection coefficient already
renormalized by the interaction at $\omega=M$. Unfortunately, these small corrections can give the contribution $\nu^{n+1}\log^n(M/\omega)$ and hence should be accounted already in the next-to-leading approximation.
The terms of different order in $\nu$ develop to small $\omega$ in different way, so substitution of $|R|^2$ by its exact value at $\omega=M$, does not help. One needs exact calculation of diagrams of  renormalized perturbation in order to include all these contributions correctly.

To summarize: in the Luttinger model with impurity the assumption of the ``poor man's'' method that conductance (aka the reflection coefficient $\Ro$) is determined by the  RG equations breaks down beyond  the one-loop approximation. Respectively, our expressions  for the physical observables (conductance and $\Ro$) differ from those obtained earlier by other authors, starting from the second loop.

\section{Matching  small reflection (transition) coefficients}

Conductance can be calculated in a different limiting case, when the coupling constant $\nu\sim 1$, but the bare reflection coefficient $|R|^2\ll 1$. For calculations in this region we use the first term in \eq{det1} and expand the exponential of the effective action in $|R|^2$. Then the integral   with respect to $\alpha$ becomes Gaussian, and we can find the Green function for the phase field and hence the effective reflection coefficient
\beq
\Ro=x\frac{1+\nu}{\pi}\int\!d\tau\frac{1-\cos\omega\tau}{|\omega|\tau^2}e^{-\sigma(\tau)},
\la{FK-ro}
\eeq
where $\sigma(\tau)$ is the result of the Gaussian integration
\beq
\sigma(\tau)=\int\!\frac{d\omega'}{2\pi}W(\omega')(1-\cos\omega'\tau).
\eeq
The expression (\ref{FK-ro}) is exact in the electron-electron interaction.

To find the asymptotic behavior of $\sigma(\tau)$ at large $\tau$  we replace $W(\omega)$ by the quadratic form for the pointlike potential in \eq{W}
and obtain
\beq
e^{-\sigma(\tau)}\approx\left(\frac{1}{{\cal M}^2(\tau^2-i\delta)}\right)^{\nu},
\la{sigma-as}
\eeq
where ${\cal M}$ is some mass determined by the properties of the electron-electron potential. Assuming that the main contribution to the integral in \eq{FK-ro} at small $\omega$   arises from
large $\tau$ we obtain \cite{Dual}
\beq
\Ro=2|R|^2(1+\nu)\Gamma(-1-2\nu)\frac{\sin\pi\nu}{\pi}\left(\frac{\omega^2+i\delta}{{\cal M}^2}\right)^{\nu}.
\la{exactnu}
\eeq

Both the  results in \eq{exactnu} and \eq{2loop-all} hold at $|R|^2\ll 1$ and $\nu \ll 1$ simultaneously. However, they are derived in the different regularization schemes and to compare them one has to  connect ${\cal M}$ with $M$. Using $W(\omega)$ in the Pauli-Villars scheme and calculating  $\sigma(\tau)$ at large $\tau$ (see \eq{sigma-as}) we obtain
\beq
{\cal M} = M e^{\gamma_E},
\la{MtoM}
\eeq
where $\gamma_E$ is the Euler constant.

Expression (\ref{FK-ro}) is valid in all loops provided that $|R|^2\ll 1$. Expanding it in $\nu$ we obtain contribution of one, two, etc. loops:
\beq
\Ro = |R|^2\left[1-\nu+\left(2+\frac{\pi^2}{6}\right)\nu^2+\ldots\right]\left(\frac{\omega}{M}\right)^{2\nu}
\la{multi}
\eeq
(three loops are given).

``Poor man's'' RG   assumes that $\Ro$ coincides with coupling constant $h_2(\omega)$. In 3 loops $h_2(\omega)$ is described by \eq{eqh2}. At small $|R|^2\ll 1$ coupling constant $h_2(\omega)$ is also small and one can substitute \eq{eqh2} by linear equation
\beq
\frac{\partial h_2}{\partial \log\omega}=2\nu h_2, \qquad \Ro = |R|^2\left(\frac{\omega}{M}\right)^{2\nu}
\eeq
Hence, this approach predicts no multi-loop corrections at $|R|^2\ll 1$. This contradicts exact expression \eq{multi}.

On the other hand, our expression correctly reproduces \eq{multi}. It can be seen from \eq{2loop-all} on the level of two loops but was also checked in all four  loops
calculated by us. The whole series in \eq{multi} is determined by renormalized perturbation theory not accounted in the ``poor man's'' RG.

In general case RG methods cannot be applied to obtain asymptotic of $\Ro$ at very small $\omega$. Renorm group is applicable in the region \eq{applicable} but the limit $\omega\to 0$ requires
that $\nu\log(M/\omega)\gg 1$. However, we know (from RG considerations) that all charges of the theory should nullify in the infrared. At a very low frequency $\omega$ one can use that all $g_{2n}(\omega)\ll 1$ and
use this smallness instead of condition $\nu\ll 1$. The leading contribution comes from the diagrams which are linear in the coupling constants $g_{2n}(\omega)$. These are the ``flower'' diagrams in Fig. 3C or Fig. 1B with an arbitrary number of ``petals''.  Contribution of these diagrams should be calculated at arbitrary $\nu$ not with the logarithmic accuracy but exactly.

Let us find also an asymptotic expression for $\Ro$ at arbitrarily low frequencies. This problem cannot be solved by an immediate application of  the RG approach since  it works only at  $\nu\ll 1$ and $\nu\log M/\omega\leq 1$. On the other hand, we know from  the RG considerations that we are dealing with the theory where all charges
nullify in the infrared. At a very low frequency $\omega$ one can use that all $g_{2n}(\omega)\ll 1$ and keep $\nu\sim 1$. The leading contribution comes from the diagrams which are linear in the coupling constants $g_{2n}(\omega)$. These are the ``flower'' diagrams in Fig. 3C or Fig. 1B with an arbitrary number of ``petals''.  Contribution of these diagrams should be calculated at arbitrary $\nu$ not with the logarithmic accuracy but exactly.

Let us note that $n$-petal diagram does not depend on bare reflection coefficient $|R|^2$ if written in terms of coupling constants $h_{2n}(\omega)$. This means that the sum of ``flower diagrams'' expressed through these coupling
constants should coincide with expansion of expression \eq{exactnu}:
\beq
\Ro=(1+\nu)\sum_{n=0}^{\infty}\frac{a_n}{n!} \nu^{n} h_{2n}(\omega), \qquad  a_n=\frac{d^{n-1}}{d\nu^{n-1}}\left[\Gamma(-1-2\nu)\frac{\sin\pi\nu}{\pi}e^{-2\nu\gamma_E}\right]_{\nu\to 0}
\la{exactnu2}
\eeq
Expression $\eq{exactnu2}$ is exact in $\nu$ provided that $\omega$ is arbitrarily small. Expansion \eq{exactnu2} accounts for all finite terms in diagrams of renormalized perturbation theory linear in coupling constants. We find $a_0=1$, $a_1=-2\ldots$, etc., in accordance with \eq{2loop-all}.

Calculation of coupling constant $h_{2n}(\omega)$ at arbitrary small $\omega$ is more complicated. In this region they are determined by linearized equation \eq{delta1}. It has a solution:
\beq
h_{2n}(\omega) = {\rm const}\left(\frac{\omega}{M}\right)^{2\nu}
\eeq
However, to obtain values of the constants one has to match this expression with the region \eq{applicable} where RG is applicable. It is possible but only perturbatively in $\nu$. In this way we find that in the limit $\omega\to 0$ \beq
\Ro = \frac{|R|^2}{|K|^2}(1-\nu-\nu\log K+ O(\nu^2))\left(\frac{\omega}{M}\right)^{2\nu}
\la{smallsmall}
\eeq
(it can be also obtained directly from \eq{final}). Let us pay attention that this expression is is singular at small $|K|^2\to 0$. In the leading order it has a pole at $K$, in the next-to-leading it contains additional $\log K$.
Moreover, it can be seen that power of logarithm is increasing in the next orders in $\nu$.

In fact, these singularities imply that at $K\to 0$ regime is changed and expression is invalid. Indeed, it is clear that for $K=0$ (impurity is impenetrable) effective reflection coefficient $\Ro=1$ at all $\omega$. As we discussed in \cite{Dual} in the narrow region $|K|^2 (M/\omega)^{2\nu}\sim 1$ (bare transition coefficient is even smaller than the frequency) there is a transition to a different regime. Reflection coefficient can be calculated as:
\beq
\Ko=|K|^2\left(\frac{M}{|\omega |}\right)^{2|\widetilde\nu|}(1+\widetilde{\nu})\frac{\sin\pi|\widetilde\nu|}{\pi}\Gamma(-1-2|\widetilde\nu|), \qquad |K|\left(\frac{M}{|\omega |}\right)^{|\widetilde\nu|}\ll 1
\label{UR}
\eeq
Here $\widetilde\nu=v_c-1$ is dual to $\nu$. Expression (\ref{UR}) has required property: at $K=0$ effective reflection coefficient $\Ro=1$.

In connection with above formulae let us discuss suggestions of the paper \cite{FRG}. It was assumed in this paper that conductance depends only on the scaling variable
\beq
\zeta=\frac{|R|}{|K|}\left(\frac{\omega}{M}\right)^{\nu}
\la{zeta-1}
\eeq
In Ref.\cite{fRG} the new approach base on so-called functional renorm group (fRG) was developed. This approach pretends to go beyond leading logarithmic approximation. Complete conductance as a function of $\zeta$ was obtained only as a result of numerical calculations. For small $\zeta\ll 1$ reflection coefficient $\Ro$ behaves as $\zeta^2$ as it should be in leading  approximation (attraction). At large $\zeta$ reflection coefficient
is close to unity and transition coefficient $\Ko$ is decreasing as some power of $\zeta$. Leading approximation predicts that $\Ko=1/\zeta^2$.
It was noted in \cite{fRG} that numerical results correspond to the {\em different} behavior of transition coefficient at large $\zeta$. Numerical curve obtained in \cite{FRG} is better fitted by
\beq
\Ko = \zeta^{2(v_c-1)} = {\rm const}\left(\frac{M}{\omega}\right)^{2|\widetilde{\nu}|}
\eeq
This power coincides with one obtained in \eq{UR}. In other words, we confirm results of \cite{fRG} regarding frequency dependence at $\zeta\gg 1$.

Beyond the leading approximation conductance is not a  function of parameter $\zeta$ only. It can be seen immediately from
our expression \eq{final} already in two-loop approximation. The reason is that in further approximations conductance is determined not by one but several coupling constants
(as it was already mentioned in \cite{fRG}) which have more complicated dependence on bare transition and reflection coefficient. Nevertheless, it seems that fRG approximation can
be far better interpolating formula (uniting two asymptotics at small and large $\zeta$) than simple one-loop approximation of Ref. \cite{M}.

\section{Conclusions}

Let us summarize. At small $\nu$ conductance in the Luttinger model with one impurity is universal. The main contribution to conductance
at the frequencies $\omega\ll a^{-1}$ comes from the large distances and only global characteristics  of the {\it ee} potential matter. Behavior of the conductance is determined by its integral strength (the Fourier component $V_{ee}=V(k\!=\!0)$) and the inverse size $ M$ which plays a role of the ultraviolet cutoff in the approximation of pointlike $ee$ potential.  Universal conductance at low frequency can be calculated for an arbitrary {\it ee} potential  and arbitrary bare reflection (transition) coefficients of the impurity with any accuracy using RG methods.  As we explained above, the well-known  ``poor man's'' method should be corrected  beyond  one loop,  but  these corrections are well under control of the standard RG methods.  The corrected expression  for the conductance smoothly matches to the expressions  obtained in the region of small reflection (transition) coefficients where conductance can be calculated exactly.

\acknowledgements
 The authors are grateful to M. Eides for very useful discussions.  The work of V. P. is supported by  the Russian Science Foundation grant \#14-22-00281.

\appendix

\section{Kubo formula}
This appendix is designed to explain the relation of \eq{obs2} to the usual Kubo formula for calculation of conductivity. Conductivity $\sigma$ is a response function to the infinitesimal electric field ${\cal E}(q,\omega)$.
\[
j(k,\omega)=\int (dq)\sigma(q,k, \omega){\cal E}(q,\omega), \qquad -i\omega \rho(k,\omega)+ik j(k,\omega)=0
\]
(owing to impurity there is no translational invariance and $\sigma$ depends on 2 momenta, $k$ and $q$).
Here $\rho$  is charge density and $j$ is the current. The last relation is due to the conservation of the current.

In the original fermion theory one can use  standard Kubo formula:
\beq
\sigma(q,k,\omega) = \frac{1}{\omega}\int dt e^{i\omega t}\langle T\{j(q,0)\; j(k,t)\}\rangle = \frac{\omega}{k q} \int dt e^{i\omega t}\langle T\{\rho(q,0)\; \rho(k,t)\}\rangle
\la{Kubo}
\eeq
where average of two currents (or charge densities) should be calculated with action based on the Hamiltonian \eq{HLuttinger} in fermion terms. If one uses a trick \cite{Hubbard} to trade electron-electron interaction
for the integration over external field $U(x,t)$:
\[
\int D\Psi D\Psi^+ \exp\left[i S_f + \frac{i}{2}\int dt dxdy\Psi^+\Psi(x)V(x-y)\Psi^+\Psi(y)\right] =
\]\[
\int DU D\Psi D\Psi^+ \exp\left[i S_f + \frac{i}{2}\int d^2x UV^{-1}U+i\int d^2x U\Psi^+\Psi(x)\right]
\]
then $\langle\rho\rho\rangle$ can be expressed as a correlator of two $U$-fields. Indeed, it is obvious that correlator of two charge densities
is equal to the average variational derivative
\beq
\langle \rho(x_1,t_1)\rho(x_2,t_2)\rangle = \int DU\exp\left[\frac{i}{2}\int d^2x UV^{-1}U\right]\frac{\delta^2}{\delta U(x_1,t_1)\delta U(x_2,t_2)}Det[U]
\eeq
where $Det[U]$ is the result of integration in fermions. Now we can take functional integrals by parts  twice and arrive at the following identity
\beq
\langle \rho(x_1,t_1)\rho(x_2,t_2)\rangle = \langle V^{-1}U(x_1t_1)V^{-1}U(x_2t_2)-V^{-1}\rangle
\la{ward}
\eeq
Relations of this type are usually called Ward identities. Hence, conductivity in the theory with interacting electrons \eq{Kubo} can be always expressed as a correlator of two $U$-fields.

Formula (\ref{obs2}) uses few further simplifications which are possible only in $d=1$ (for details see \cite{Dual}). Here one can use the conservation of axial current in the external field $U$
\beq
\partial_t j +\partial_x\rho = \frac{1}{\pi}\partial_x U + {\mathfrak D}(t)\delta(x)
\eeq
The first term in the right hand side of this equation is the Adler anomaly, the second describes non-conservation of the axial current owing to the impurity.
Here ${\mathfrak D}(t)$ is the charge jump at $x=0$. Together with conservation of the vector current
the last  equation allows  to determine completely $\rho(x,t)$ and $j(x,t)$ in the external field. The addition to the ballistic current appears to be proportional to
${\mathfrak D}(t)$. The last quantity depends only on the electron phase $\alpha(t)$ at the point of impurity. Integrating in fermions and external field $U(x,t)$
at fixed $\alpha$ we arrive at the effective action consisting of two terms, \eq{kin-energy} and \eq{det1}.

To obtain the conductivity one has to place the system in the external electric field and differentiate the current in this field. In the gauge $A_1=0$ additional electric field
is a shift $U\to U+\varphi$ and the addition to the ballistic current is proportional to variational derivative
 \begin{equation}
\delta j[U](x,t)=\int d^2x'e^{-\frac{1}{2}UV^{-1}U}\varphi (x')\frac{\delta}{\delta U(x')}[j[U](x){\mathfrak Det}[U]]
\label{previouspaper}
\end{equation}
After transition to electron phase $\alpha$ and integration in all $U$ we obtain \cite{Dual} (electrical field ${\cal E}=-\partial_x\varphi$)
 \begin{equation}
\delta j(\omega,k)= \frac{2\omega |\omega |}{\omega^2- v_c^2k^2+i\delta}\frac{v_c}{\pi}|{\cal R}_{\omega}|^2\int\frac {dq}{(2\pi)} \frac{{\cal E}(q,\omega)}{\omega^2- v_c^2q^2+i\delta}
\label{jf}
\end{equation}
It depends only on effective reflection coefficient $|R_\omega|^2=1-|K_\omega|^2$ equal to
\beq
|R_\omega|^22\pi\delta(\omega-\omega')= \frac{i\pi}{|\omega|W(\omega)v_c}\frac{1}{{\cal Z}}\int D\alpha\; \alpha(-\omega'){\mathfrak D}(\omega)e^{-S_{kin}[\alpha]}{\mathfrak Det}_{imp}[\alpha]
\eeq
Expression for $|R_\omega|^2$ in this form was used in \cite{Dual}. It is obtained from \eq{previouspaper} by integtation by parts in $\alpha$.
One can use that the charge jump is a derivative of ${\mathfrak Det}_{imp}[\alpha]$
\beq
{\mathfrak D}(\omega)=2i\frac{\delta}{\delta\alpha(-\omega)}\log{\mathfrak Det}_{imp}[\alpha]
\eeq
and integrate by parts again. We arrive at
\beq
|R_\omega|^2= \frac{2\pi}{|\omega|W^2(\omega)v_c}\frac{1}{{\cal Z}}\int D\alpha\; [W(\omega)\delta (\omega-\omega') - \alpha(-\omega')\alpha(\omega)]e^{-S_{kin}[\alpha]}{\mathfrak Det}_{imp}[\alpha],
\label{jf1}
\eeq
The first term in the square brackets is the Green function $G_0(\omega)$ in the free theory without impurity, the second is a Green function $G(\omega)$. Therefore, \eq{jf} coincides with \eq{obs2} of the main text. Expression for conductance \eq{obs14} can be derived from \eq{jf} in the limit of constant electrical field.

\section{Regularization scheme dependence}

Let us demonstrate that Gell-Mann-Low equations for coupling constant $h_2$ depends on the regularization scheme already on the level of 2 loops. We mentioned that in $\nu^2$-order diagram IB does not contribute to this equation. This is true only in Pauli-Villarse regularization \eq{Gpv}. In other regularizations
\beq
\frac{\partial h_2}{\partial\log\mu}= \frac{2\nu}{1+\nu h_2} h_2(1-h_2)+c_1\nu^2 h_2(1-h_2)(1-2h_2).
\la{h2-all}
\eeq
The last term describes contribution of this diagram and $c_1$ depends on the regularization. This contribution is proportional to the coupling constant $h_6$ which can be found from \eq{solh}. Iterating \eq{h2-all} we obtain the following addition to \eq{h4} for $g_2(\mu)$:
\beq
\delta g_2(\mu)=\frac{\nu c_1\K(M/\mu)^{2\nu}}{[\R+\K(M/\mu)^{2\nu}]
^2}.
\la{h2-add}
\eeq

One can come to the same conclusion without direct calculation of diagram IB. Indeed, coupling constants in different regularization schemes are related by change in the ultraviolet cutoff $M$. Let us assume that in the new regularization cutoff $\widetilde{M}=Me^{c_2}$ with $c_2\ll 1$. Expanding coupling constant $\widetilde{g}_2(\mu)$ in the new regularization
in $c_2$ and using \eq{h4}, we will obtain $\delta g_2(\mu)=\widetilde{g}_2(\mu)-g_2(\mu)$  coinciding with \eq{h2-add} ($c_1=-2c_2$).

Thus change of the regularization scheme change Gell-Mann-Low equations in the second loop. For this reason physical conductance {\em cannot} be determined from any RG equation in
this approximation. Some additional terms which can cancel the scheme dependence are necessary. These terms also depends on the regularization scheme (terms proportional to $g_4(\omega)$ in \eq{2loop-all})in such a way that physical observable $\Ro$ does not depend on the scheme.

Usually, dependence of the Gell-Mann-Low function on reqularization starts later, only in the third loop. This is related to the specific behavior of $\beta$-function. Indeed,
let us consider standard Gell-Mann-Low function $\beta(g)=b_1g^3+b_2g^5+b_3g^5+\ldots$. We present Gell-Mann-Low equation in the form
\[
-\frac{\partial (1/g^2)}{\partial\log\mu}=2b_1+
2b_2/g^{-2} +2b_3/g^{-4},
\]
Solution of this equation at $\log{(M/\mu)}\gg 1$ is
\[
\frac{1}{2g^2}=  b_1\log \frac{M}{\mu} + b_2 \log\log\frac{M}{\mu}+\frac{b_3}{\log\frac{M}{\mu}}+\ldots
\]
Change of the regularization scheme $M\to\widetilde{M}=Me^{c_2}$ change Gell-Mann-Low coefficients
\[
\frac{1}{2g^2}=  b_1\log \frac{M}{\mu} + b_2 \log\log\frac{M}{\mu}+\frac{b_3+c_2b_2}{\log\frac{M}{\mu}}+\ldots
\]
but only in the third loop.

\end{document}